\documentclass[english,3p,final]{elsarticle}
\usepackage{lmodern}
\usepackage{lmodern}
\usepackage[T1]{fontenc}
\usepackage[latin9]{inputenc}
\usepackage{babel}
\usepackage{float}
\usepackage{mathtools}
\usepackage{amsmath}
\usepackage{graphicx}
\usepackage{esint}
\PassOptionsToPackage{normalem}{ulem}
\usepackage{ulem}
\usepackage[unicode=true,
 bookmarks=false,
 breaklinks=false,pdfborder={0 0 1},backref=false,colorlinks=false]
 {hyperref}
\usepackage{color}

\makeatletter

\providecommand{\tabularnewline}{\\}

\usepackage{babel}

\makeatother

\begin{document}
\begin{frontmatter}

\title{\textbf{Dynamic mesh refinement for discrete models of jet electro-hydrodynamics}}

\author[iac]{Marco Lauricella}

\author[iac]{Giuseppe Pontrelli}

\author[salento,nano]{Dario Pisignano}

\author[iac]{Sauro Succi\corref{cor1}}

\ead{s.succi@iac.cnr.it }

\cortext[cor1]{Corresponding author}

\address[iac]{Istituto per le Applicazioni del Calcolo CNR, Via dei Taurini 19,
00185 Rome, Italy}

\address[salento]{Dipartimento di Matematica e Fisica \textquotedbl{}Ennio De Giorgi\textquotedbl{},
University of Salento, via Arnesano, 73100 Lecce, Italy}

\address[nano]{Istituto Nanoscienze-CNR, Euromediterranean Center for Nanomaterial
Modelling and Technology (ECMT), via Arnesano, 73100 Lecce, Italy}

\begin{abstract}
Nowadays, several models of unidimensional fluid jets exploit discrete
element methods. In some cases, as for models aiming at describing the electrospinning nanofabrication process of polymer fibers, discrete
element methods suffer a non-constant resolution of the jet representation.
We develop a dynamic mesh-refinement method for the numerical study
of the electro-hydrodynamic behavior of charged jets using discrete
element methods. To this purpose, we import ideas and techniques from
the string method originally developed in the framework of free-energy
landscape simulations. The mesh-refined discrete element method is
demonstrated for the case of electrospinning applications. 
\end{abstract}
\end{frontmatter}

\section{Introduction}

Discrete element methods are widely employed to model fluid flows
in air channels, pipes and several other applications, such as modeling
ink-jet printing processes, electrospinning, spray jets, micro-fluid dynamics
in nozzles, etc. \cite{yarin2001taylor,feng2002general,apostolou2008discrete,wijshoff2008structure,dorr2008spray}.
In particular, unidimensional jets can be easily modeled as a sequence
of discrete elements by defining a mesh of points, which is used to
discretize a continuous object (e.g. a liquid body) as a finite sequence
of discrete elements. The aim of such model is to provide a relatively
simple computational framework based on particle-like ordinary differential
equations, rather than on the discretization of the partial differential
equations of continuum fluids. Electro-hydrodynamic flows, however,
are often subject to strong interactions leading to major deformation
of the jet, hence to significant heterogeneities in the spatial
distribution of the discrete particle. The latter, in turn, imply
a loss of accuracy of the numerical method, since the most stretched
portions of the jet become highly under-resolved. One of these cases
is the electrospinning process, where a polymeric liquid jet
is ejected from a nozzle and accelerated towards a conductive collector
by a strong electric field. In this framework, the jet is stretched
so that its diameter decreases below the micrometer-scale, providing
a one-dimensional structure with very high surface area to volume ratio. This intriguing
feature of the resulting polymer nanofibers spawned several papers \cite{reneker1996nanometre,li2004electrospinning,carroll2008nanofibers,luo2012electrospinning,persano2013industrial,montinaro2015dynamics}
and books \cite{ramakrishna2005introduction,pisignanoelectrospinning,wendorff2012electrospinning}
focussing on the electrospinning process.

In this framework, pioneering works by the Reneker and Yarin groups were
focused on developing \textit{ad-hoc} discrete element methods for
electrospinning, which describe nanofibers as a series of
beads obeying the equations of Newtonian mechanics \cite{yarin2001taylor,reneker2000bending}.
This modeling approach has gained an important role in predicting the outcome of
electrospinning experiments. In addition, such models might support experimental
researchers with a likely starting point for calibrating processes
in order to save time before subsequent optimization work.

In these models, the excess charge is distributed to each
element composing the jet representation, and it is static in the
frame of reference of the extruded fluid jet. Once the solution surface
tension is overcome by electrical forces at the spinneret, the jet
serves as fluid medium to push away the mutually repulsing electric
charges from the droplet pending at the nozzle of the apparatus. The
main forces affecting the jet dynamics, which are accounted for in
such models, include viscoelasticity, surface tension and
electrostatic interactions with the external field and among excess
charges in the liquid. In particular, the Coulomb repulsion between
electric charges triggers bending instabilities in electrically charged
jets, as demonstrated by Reneker et al. \cite{reneker2000bending}.
Despite being often neglected, air drag and aerodynamic effects, which
may also lead to bending instabilities, have been recently modeled
by discrete element methods \cite{lauricella2016three,Lauricella2015dissipative,lauricella2015nonlinear}.
Further, several complex viscoelastic models were included in order to simulate viscoelastic Boger fluid solutions \cite{pontrelli2014electrospinning,carroll2006electrospinning}.
Systematic investigations were carried out on several simulations
parameters: polymer concentration, solution density, electric potential,
and perturbation frequency \cite{coluzza2014ultrathin,thompson2007effects}.

Notwithstanding a satisfactory, though generally qualitative, agreement
with experimental results has been shown \cite{lauricella2015jetspin,carroll2011discretized,reneker2000bending}, 
all these methods suffer the non-constant resolution of the jet representation, which usually
decreases downstream. Indeed, given a uniform discretization of the
initial polymeric drop pending at the nozzle, the initial step of
the jet dynamics is characterized by the prevalent viscoelastic force,
and it leads to a second regime, when the longitudinal stress changes
under the effect of the applied electric voltage, accelerating the
jet towards the collector \cite{reneker2000bending}. Hence, a free-fall
regime describes the later jet dynamics. Since the longitudinal viscoelastic
stress along the fiber is usually larger close to the nozzle, the
distance between each discrete element increases as a result of the
uniformly accelerated motion, which drives the farthest elements from
the spinneret. As a consequence, the jet discretization close the
collector becomes rather coarse to model efficiently the filament,
and the information (position, velocity, radius, stress, etc.) describing
the jet is lost downstream.

A refined description of jet is necessary in every instance where
a high-resolution description of physical quantities is requested,
such as for bending instabilities, varicose instabilities of diameter,
etc. Further, a dense mesh provides a more strict assessment of the
Coulomb repulsion term, which is important to properly account the
transverse force acting on the jet. Recently, refinement procedures
were proposed in few works in order to avoid low resolution problems
downstream \cite{kowalewski2009modeling,kowalewski2005experiments}.
Nonetheless, these procedures exploit only linear interpolations,
and are not able to impose a uniform mesh of jet representation, which
is likely the simplest choice to model properly the entire jet, from
the nozzle to the collector.

Here, we present an algorithm specifically developed to address the
issue. The aim of the algorithm is to recover a finer jet representation
at constant time interval, before the information describing the jet
is scattered downstream, so as to preserve the jet modeling representation
and make the simulation more realistic. Further, the algorithm can
be used to enforce several types of mesh depending on the physical
quantities under investigation, providing an adaptive description
of the process.

The article is organized as follows. 
In Sec. 2, we summarize the 3D model for electrospinning, with the set of equations of motion (EOM) which govern the dynamics of system. 
In Sec. 3, we present the algorithm,
with the relative step for its implementation. In Sec. 4 we report
a numerical example of its application within a discrete element method
for electrospinning modeling. Finally, conclusions are outlined in
Sec. 5.

\section{Model\label{sec:The-model}}

In this work, we apply our algorithm to the electrospinning model implemented in JETSPIN, an open-source code
specifically developed for electrospinning simulation \cite{lauricella2015jetspin}, which is briefly summarized in the following text. The model is an extension of the discrete model originally introduced by Reneker et al. \cite{reneker2000bending}. The main assumption of such model is that the filament can be reasonable represented by a series of $n$ discrete elements (jet beads). Each one of this elements is a particle-like labelled by an index $i-th$. A Cartesian coordinate system is taken so that the origin is located at the nozzle, and $x-axis$ is pointing perpendicularly towards a collector where the nanofiber is deposited. The nozzle is also represented as a particle like (nozzle bead) which can moves on the plane $y-z$ (further details in the following text). Given a fluid jet starting at the nozzle and labeling $i=0$ the nozzle bead, we obtain the set of beads indexed $i=0,1,...,n$, where $n$ denotes the bead of the other extremity of the filament. Thus, given $\vec{r}_i$ the position vector of a generic $i-th$ bead, we define the mutual distance between $i$ and $ i+1$ elements 
\begin{equation}
l_{i}=\left|\vec{\textbf{r}}_{i+1}-\vec{\textbf{r}}_{i}\right|.
\label{eq:define-l}
\end{equation}
Here, $l_i$ stands for the length step used for discretizing the filament. Note that the length $ l$
is typically larger than the filament radius, but smaller than the characteristic lengths of other observables of interest (e.g. jet curvature radius). Denoted  $m_{i}$ the mass,
$t$ the time, and $\vec{\boldsymbol{\upsilon}}_{i}$ the velocity vector of $i-th$ bead, its acceleration is given by Newton's law
\begin{equation}
\frac{d\vec{\boldsymbol{\upsilon}}_{i}}{dt}=\frac{\vec{\textbf{f}}_{tot,i}}{m_{i}}\:\text{,}
\label{eq:force-EOM}
\end{equation}
where $\vec{\textbf{f}}_{tot,i}$ is the total force acting on $i-th$ bead, given
as a sum of several terms
\begin{equation}
\vec{\textbf{f}}_{tot,i}=\vec{\textbf{f}}_{el,i}+\vec{\textbf{f}}_{c,i}+\vec{\textbf{f}}_{\upsilon e,i}+\vec{\textbf{f}}_{st,i} \:\text{.}\label{eq:force-sum}
\end{equation}

In last Eq, $\vec{\textbf{f}}_{el,i}$ stands for the
electric force due to the external electrical potential $\Phi_{0}$ imposed between the nozzle and the 
collector located at distance $h$ along versor $\vec{\textbf{x}}$, and given as
\begin{equation}
\vec{\textbf{f}}_{el,i}=q_{i}\frac{\Phi_{0}}{h}\cdot\vec{\textbf{x}}\text{.}
\end{equation}
Denoted $q_{i}$ the charge of $i-th$ bead, 
the net Coulomb force $\vec{\textbf{f}}_{c}$ is
\begin{equation}
\vec{\textbf{f}}_{c,i}=\sum_{\substack{j=1\\
j\neq i
}
}^{n}\frac{q_{i}q_{j}}{\left |\vec{\textbf{r}}_{j}-\vec{\textbf{r}}_{i}  \right |^{2}}\cdot\vec{\textbf{u}}_{ij}
\label{eq:force-coul}
\end{equation}
acting on the $i-th$
bead and originated from all the other $j-th$ beads. Further, we denote $\vec{\textbf{f}}_{st}$ the surface tension force 
\begin{equation}
\vec{\textbf{f}}_{st,i}=k_{i}\cdot\pi\left(\frac{a_{i}+a_{i-1}}{2}\right)^{2}\alpha\cdot\vec{\textbf{c}}_{i}\text{,}
\end{equation}
acting on $i-th$ bead towards the center of the local curvature 
(according to $\vec{\textbf{c}}_{i}$ versor)
to restore the rectilinear shape of the jet, 
given $\alpha$ the surface tension coefficient, $a_{i}$ the jet radius, and $k_{i}$ the local curvature
(measured at $i-th$ bead).
Finally, $\vec{\textbf{f}}_{\upsilon e}$ stands for the viscoelastic force
\begin{equation}
\vec{\textbf{f}}_{\upsilon e,i}=-\pi a_{i}^{2}\sigma_{i}\cdot\vec{\textbf{t}}_{i}+\pi a_{i+1}^{2}\sigma_{i+1}\cdot\vec{\textbf{t}}_{i+1}\text{,}
\label{eq:viscoelastic-force}
\end{equation}
pulling bead $i$ back to $i-1$ and towards $i+1$,
with $\sigma$ the stress and $\vec{\textbf{t}}_{i}$ the unit vector pointing bead $i$
from bead $i-1$.
The polymeric jet is assumed as a viscoelastic Maxwellian liquid so that
$\sigma$ evolves in time following the constitutive equation:

\begin{equation}
\frac{d\sigma_{i}}{dt}=\frac{G}{l_{i}}\frac{dl_{i}}{dt}-\frac{G}{\mu}\sigma_{i},\label{eq:stress-ode}
\end{equation}
where $G$ is the elastic modulus, and $\mu$ is the viscosity
of the fluid jet.

Finally, the velocity $\vec{\boldsymbol{\upsilon}}_{i}$ satisfies the
kinematic relation:

\begin{equation}
\frac{d\vec{\textbf{r}}_{i}}{dt}=\vec{\boldsymbol{\upsilon}}_{i} . \label{eq:pos-EOM}
\end{equation}
The set of three Eqs \ref{eq:force-EOM}, \ref{eq:stress-ode} and \ref{eq:pos-EOM}
governs the time evolution of system.

The nozzle bead with
charge $\bar{q_0}$ and $x_0=0$ is experiencing uniform circular motion 
in order to model fast mechanical oscillations of the spinneret\cite{coluzza2014ultrathin}, 
where we denote $A$ the amplitude of the perturbation and $\omega$ its frequency.

The initial simulation conditions as well as the jet insertion at the nozzle are briefly described in the following.
We assume that all the electrospinning simulations start with
a single jet bead, which is placed at distance $l_{step}$ from the
nozzle along the $x$ axis. The single bead represents a jet with a cross-sectional
radius $a_{0}$, defined as the radius of the filament at the nozzle
before the stretching process. The jet bead has an initial velocity
$\upsilon_{s}$ along the $x$ axis which is impressed by the bulk
fluid velocity in the syringe needle. Evolving in time the system,
this jet bead travels up to a distance $2\cdot l_{step}$ away from
the nozzle, and a new jet bead is placed at distance
$l_{step}$ between the two previous bodies (nozzle and previous jet
bead). The procedure is repeated so that we obtain a series of $n$
beads representing the jet. Therefore, the parameter $l_{step}$
represents the length step which is used to discretize the polymer solution filament
in discrete elements, and, in other words, it is the spatial resolution imposed at the nozzle before the stretching process
starts acting.

Each simulation starts with only
two bodies: the nozzle bead at fixed at $x_0=0$ and a single bead modelling 
an initial jet segment of cross-sectional radius $a_{init}$ located at distance $l_{step}$ from
the nozzle along the $x$ axis with an initial
velocity $\upsilon_{b}$ along the $x$ axis equal to the bulk fluid
velocity in the nozzle. Here, $l_{step}$ denotes the length
step used to discretize the polymer solution filament
in discrete elements, and, in other words, it is the spatial 
resolution imposed at the nozzle before the stretching process
starts acting.
Under the effect of the external electric field the jet bead starts
to move towards the collector, and as soon as it is a
distance of $2\cdot l_{step}$ away from the nozzle, a new jet bead
(third body) is placed at distance $l_{step}$ from the nozzle along
the straight line joining the two previous bodies (further details in Ref. \cite{lauricella2015jetspin}).

\section{Algorithm\label{sec:The-algorithm}}

\begin{figure}[H]
\begin{centering}
\includegraphics{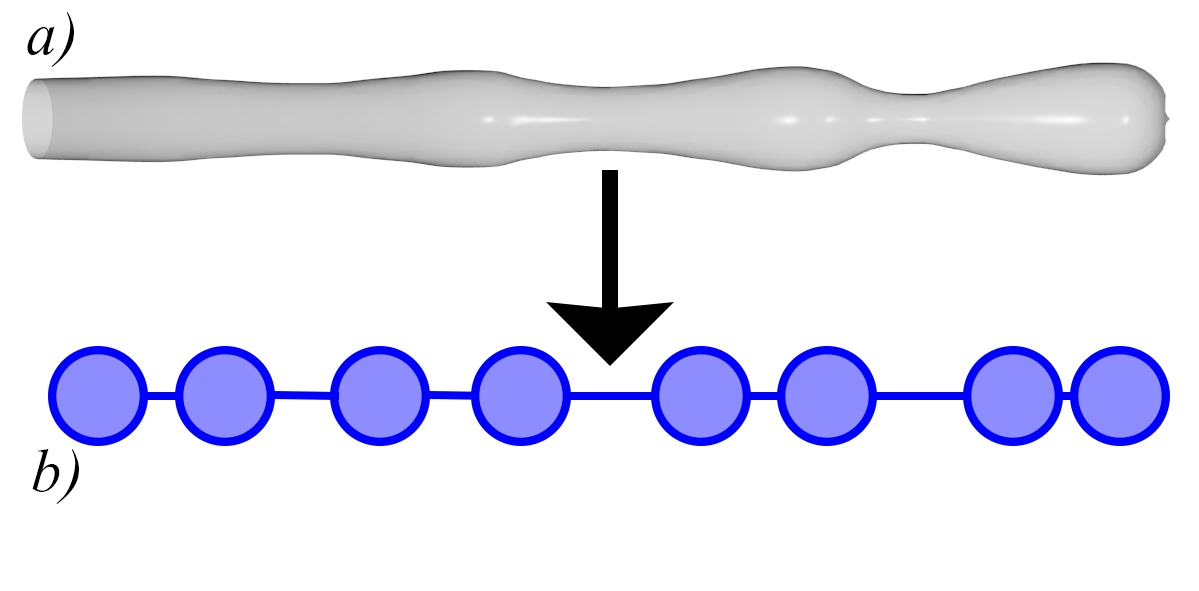} 
\par\end{centering}

\protect\caption{Sketch of a liquid jet (a) and its modeling representation by discrete
elements (b) drawn as circles with connections representing
the mutual interactions between two consecutive elements.}

\label{Fig:DEM-diagram} 
\end{figure}

Given a series of discrete elements representing a liquid jet (Fig. \ref{Fig:DEM-diagram}), our main objective is to recover a finer
representation, as soon as it has become too coarse. 
As example, if our aim is to investigate 
cross section fluctuations of the nanofiber deposited on the collector at
a length scale of 1 mm, as soon as the stretching process moves two beads
of the discretized jet  at mutual distance larger than 1 mm, the representation is too coarse for resolving
such fluctuations at the desired length scale.
For this purpose,
we note that the jet elements define a natural parametrization of
a string (curve), which starts and ends at two given extreme points.
Thus, we borrow and adapt ideas and techniques from elements of the
simplified string method originally proposed by E. Weinan et al. \cite{weinan2007simplified}
for computing the minimum energy path in an energy landscape context.
In the original simplified string method, a string is dynamically evolved in the free energy landscape
in order to identify the reactive path connecting two minima through a saddle point (usually the transition state of the
process under investigation). An initial string is discretized by a uniform mesh (points equally spaced). 
Then, the string (each point of the mesh) is
evolved in time by a steepest descent dynamics and the mesh undergoes the break of its uniform distribution.
At constant time interval, a check on the mesh uniformity is performed, and, if necessary, 
the uniform representation of the mesh is reinforced by a cubic spline interpolation.
Here, our idea is to apply a refinement procedure at constant time interval
during the jet hydrodynamics whenever the mutual distance between
any two consecutive elements representing the jet exceeds a prescribed
threshold. Note that the refinement procedure should be performed
before the jet representation is scattered by the stretching dynamics.
In other words, we wish to preserve the modeling information before
it vanishes.

\begin{figure}[H]
\begin{centering}
\includegraphics[scale=0.5]{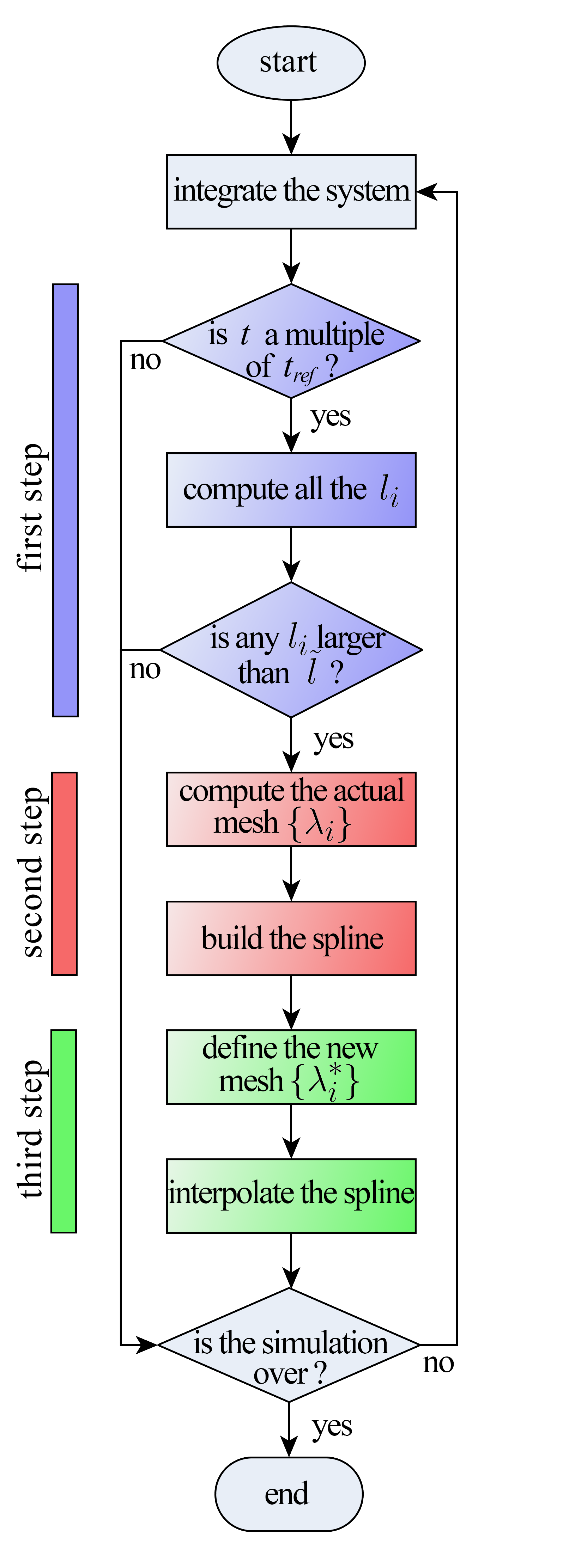} 
\par\end{centering}

\protect\caption{Flow chart of the algorithm. The three steps of the procedure are
highlighted in different colors.}

\label{Fig:flow} 
\end{figure}

By integrating in time the equations of motion above 
(see the field "integrate the system" in Fig \ref{Fig:flow}),
the observable $l_{i}$
defined in Eq \ref{eq:define-l}
between two consecutive elements $i$ and $i+1$  is found to increase under
the main action of the external electric field. 
The algorithm proceeds in three steps as follows:

\noindent \textit{\uline{First step:}}\textit{ }

\begin{figure}[H]
\begin{centering}
\includegraphics[scale=0.5]{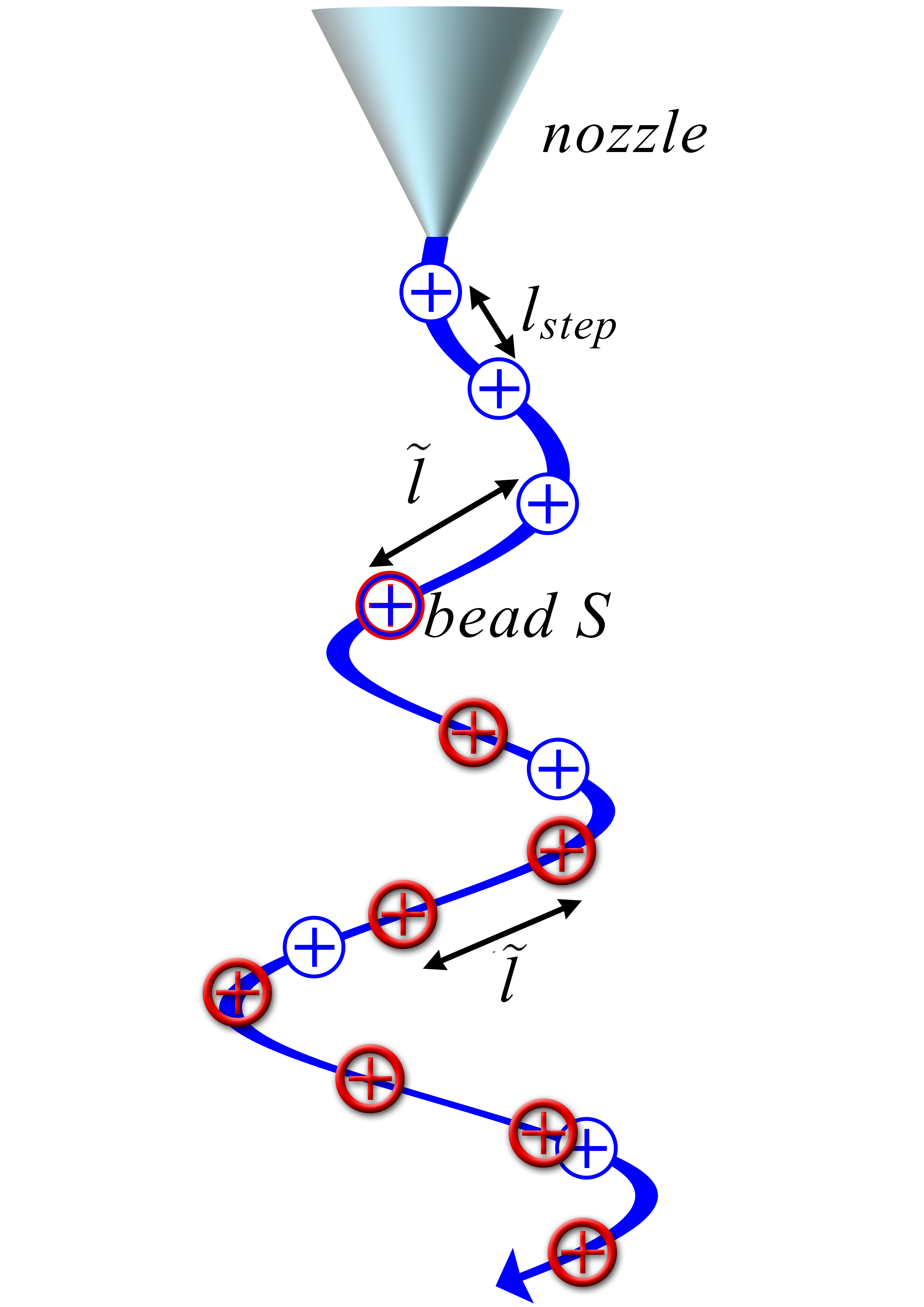} 
\par\end{centering}

\protect\caption{Diagram (not in scale) showing the actual jet representation (in blue),
and the refined representation with uniform discretization of arc-length
equal to $\tilde{l}$ (in red) starting from the bead with index $s$.
We report the length step $l_{step}$ used to discretize the jet at
the nozzle before the stretching. We also highlight the threshold
distance $\tilde{l}$ used to discern the bead where the spline interpolation
starts (bead with index $s$). Note the string is interpolated by
using a constant arc-length equal to $\tilde{l}$.}

\label{Fig:rep-interp} 
\end{figure}

At regular interval time $t_{ref}$, we compute the mutual distance
$l_{i}$ between all the elements of the actual jet representation.
Note that
the parameter $t_{ref}$ should be set in order to perform the check
before the information of the jet representation is lost due to the stretching.
Starting from the nozzle $(i=0)$, all the mutual distance $l_{i}$
are tested for checking whether the length $l_{i}$ is larger than
a prescribed threshold length $\tilde{l}$. In particular, we mark
by $s \in [0,...,n]$ the bead closest to the nozzle, whose distance $l_{s}$ to
the next bead $s+1$ is larger than threshold $\tilde{l}$ (see Fig
\ref{Fig:rep-interp}). If $l_{i}<\tilde{l}$ for any $i$, the simulation
proceeds normally, otherwise we move to the second step.

\noindent \textit{\uline{Second step:}}

We introduce the curvilinear coordinate $\lambda\in[0,1]$ to parametrize
the jet path, where $\lambda=0$ identifies the nozzle, and $\lambda=1$
the jet at the collector. For each bead we compute its respective
$\lambda_{i}$ value by using the formula:

\begin{equation}
\lambda_{i}=\frac{1}{l_{path}}\sum_{k=1}^{i}l_{k}\label{eq:first}
\end{equation}
where $l_{path}=\sum_{k=1}^{n}l_{k}$ is the jet path length from the nozzle
to the collector. In other words, we calculate the arc-lengths corresponding
to the current parametrization of the string (polymer solution jet). The set
of ${\lambda_{i}}$ values represent the mesh used to build a cubic
spline. Denoted generically by $y$ one of the main quantities describing
the jet, the data $y_{i}=y\left(\lambda_{i}\right)$ are tabulated
and used in order to compute the coefficients of the spline. Note
that several piecewise cubic functions are available in the literature \cite{de2001practical}.
Here, we use the Akima's algorithm \cite{akima1970new} in order to compute
spline coefficients, since it was demonstrated that other natural
cubic splines could oscillate, whenever the tabulated values change
quickly \cite{de2001practical}. The cubic coefficients are computed
for all the main quantities describing the jet beads (positions, fiber
radius, stress, velocities)

\noindent \textit{\uline{Third step:}}

The filament is refined starting from the bead with index $s$ up
to the last bead (the farthest bead from the nozzle $i=n$), and only
in this part of the string a uniform discretization of arc-length
equal to $\tilde{l}$ is imposed to the representation. In fact, we use a uniform discretization only for the sake of simplicity, and
this choice is not mandatory. The interpolation is performed by using
the spline coefficients previously computed in the \textit{second
step}. Starting from the bead with index $s$, we enforce the equal
arc-length parametrization (uniform parametrization) of the filament
by imposing all the arc-lengths of the elements to be equal to $\tilde{l}$
(see Fig \ref{Fig:rep-interp}). First of all, we define the number
of beads in the new representation as:

\begin{equation}
n^{*}={\displaystyle s+\left\lceil \frac{\sum_{k=s+1}^{n}l_{k}}{\tilde{l}}\right\rceil },
\end{equation}
where the symbol $\left\lceil \cdot\right\rceil $ denotes the ceiling
function. Denoted $n_{ref}=\left\lceil \frac{\sum_{k=s+1}^{n}l_{k}}{\tilde{l}}\right\rceil $,
the new mesh ${\lambda_{i}^{*}}$ is obtained as:

\begin{equation}
\lambda_{i}^{*}=\begin{cases}
\lambda_{i} & \text{if}\,i\leq s\\
\lambda_{i-1}^{*}+\frac{1-\lambda_{s}}{n_{ref}} & \text{if}\,i>s
\end{cases}
\end{equation}
where $\lambda_{s}$ is the value of the parameter $\lambda$ at the
bead $s$. Note that it is still $\lambda^{*}\in[0,1]$, and the parameter
$\lambda^{*}$ has the same values from zero (the nozzle) up to the
bead $s$. In this way, we do not apply the refinement procedure close
to the nozzle, where new beads are added to the system, in order
to not perturb the mechanism of bead injection (for the injection
mechanism see Ref \cite{lauricella2015jetspin}). Finally, the new
values $y\left(\lambda_{i}^{*}\right)$ are computed for all the $\lambda_{i}^{*}$
by \textit{spline interpolation}. The procedure is performed for $y$ to be positions,
fiber radius, stress and velocities of the jet beads in order to provide
the corresponding quantities in the new mesh ${\lambda_{i}^{*}}$. Then,
the new system is evolved in time following Eqs. 1. It is worth stressing
that, in principle, it is also possible to define an adaptive mesh
of the string by using a suitable weight function of a generic observable
$O$ of the string (e.g. jet radius).

A summarizing flow chart of the algorithm has been sketched in Fig
\ref{Fig:flow}. It is worth stressing that, notwithstanding the length
of the curve in Eq \ref{eq:first} is computed by linear interpolation,
the above procedure of reparametrization has the accuracy of a third-degree
polynomial, since we use Akima's cubic spline for the interpolation
of the curve. This is fine, since we care mostly about preserving
the string accuracy than computing carefully its arc-length.

\section{Applications on electrospinning}

The dynamic refinement method was implemented in a modified version
of JETSPIN, an open-source code specifically developed for electrospinning
simulations \cite{lauricella2015jetspin}. The JETSPIN
code delivers a discrete element model, which bases on the model originally introduced by Reneker et al. \cite{reneker2000bending}.
As a test case, we perform a simulation for a polyvinylpyrrolidone
(PVP) solution, which is commonly used in electrospinning experiments
\cite{persano2013industrial,reneker2008electrospinning}. The simulations
are carried out by using the simulation parameters reported in Ref.
\cite{lauricella2015jetspin}, modeled on the experimental
data reported by Montinaro et al. \cite{montinaro2015dynamics}. For
convenience, all the simulation parameters are summarized in Tab. \ref{tab:simulation-param}.

\begin{figure}[H]
\begin{centering}
\includegraphics[scale=0.5]{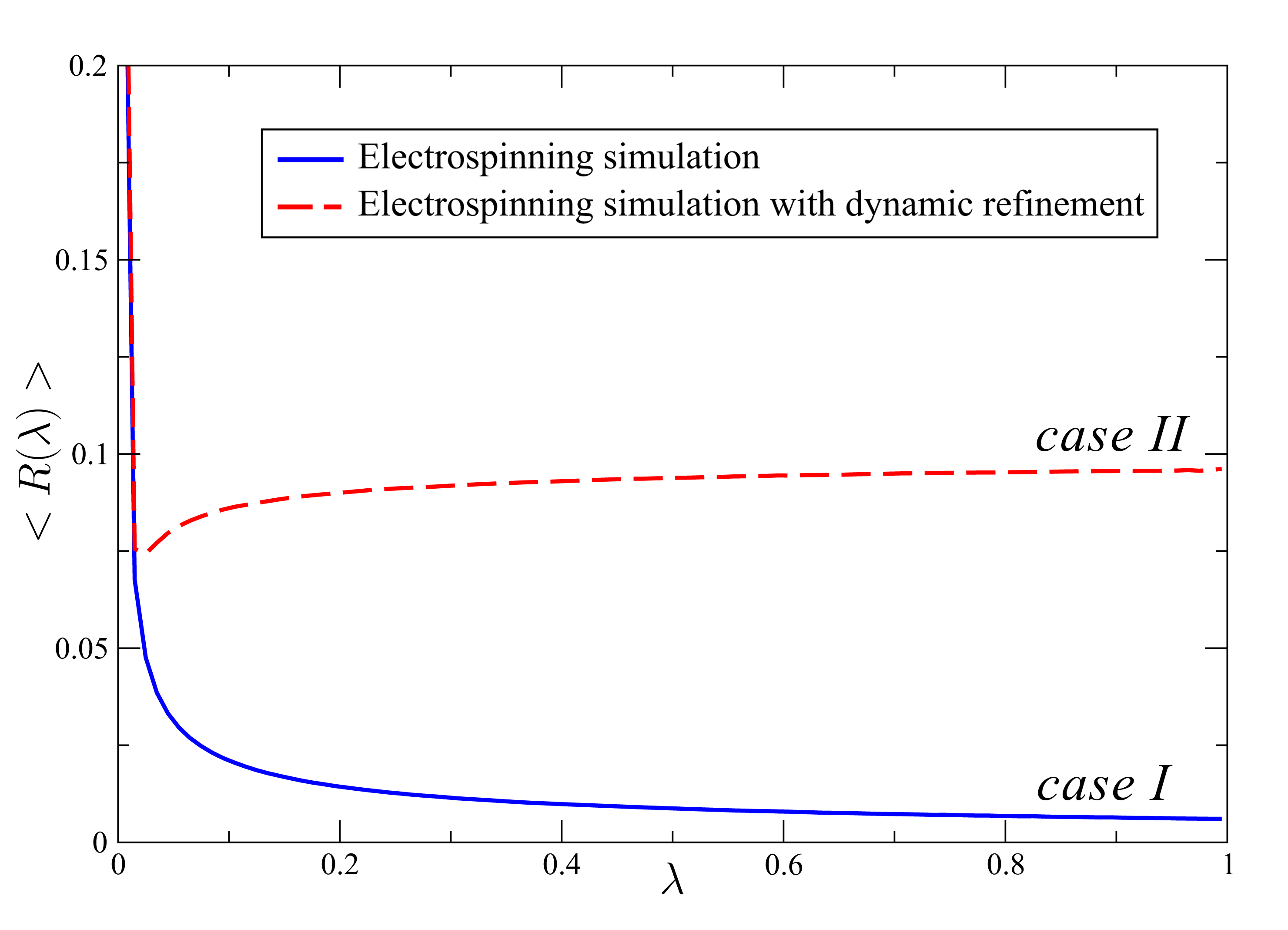} 
\par\end{centering}

\protect\caption{Mean value of the resolving power $<R\left(\lambda\right)>$ computed
as function of the curvilinear coordinate $\lambda$ over the simulation
time in stationary regime for the \textit{case I} (blue line) and the
\textit{case II} (red dashed line) introduced in the text, respectively.}

\label{Fig:resolving} 
\end{figure}

\begin{figure}[H]
\begin{centering}
\includegraphics[scale=0.5]{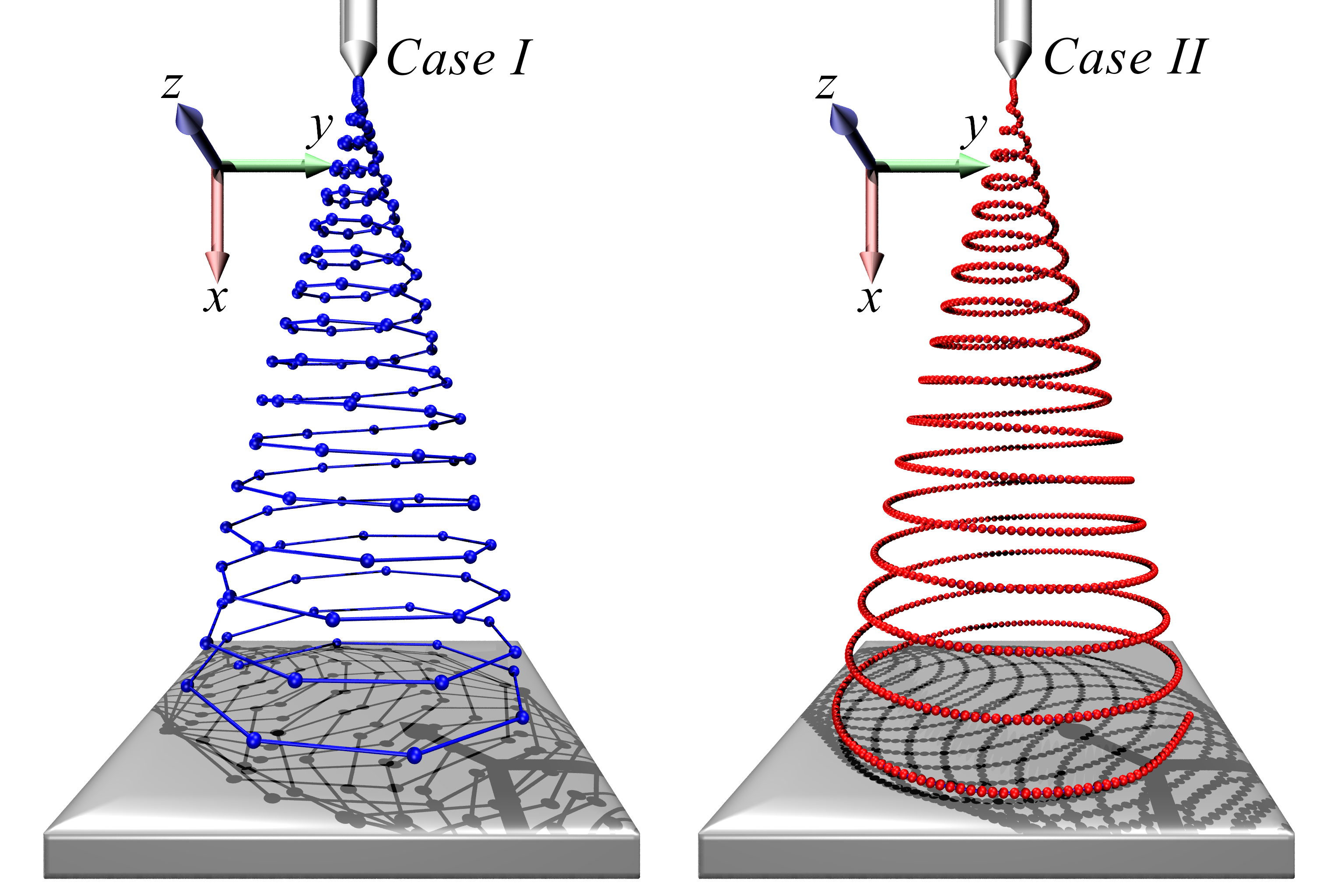} 
\par\end{centering}

\protect\caption{Two snapshots showing the jet representation in the \textit{case I}
(in blue, left panel), and in the \textit{case II} (in red, right
panel). The refined procedure was carried out in the \textit{case
II} imposing a uniform discretization with arc-length $\tilde{l}$ equal to 0.2 
cm.}

\label{Fig:representation-filament} 
\end{figure}

We now consider two different simulations (in the following \textit{case
I} and \textit{case II}), the second incorporating the mesh-refined
algorithm. The simulations were integrated in time for 0.2 second
by the fourth order Runge\textendash Kutta method with a time step
equal to $10^{-8}$ sec, and the jet beads were inserted at a distance
$l_{step}=0.02$ cm from the nozzle. In the second simulation the
dynamic refinement is applied every $t_{ref}=0.001$ seconds.
During the refinement procedure we impose all the arc-lengths $\tilde{l}$
equal to 0.2 cm for the new parametrization. In both simulations,
we distinguish two different regimes of the observables describing
the electrospinning process. In the first, the observables follow
an initial drift when the jet has not reached the collector yet.
Then, all the observables fluctuate around a constant mean value (stationary
regime) after the jet touches the collector. Here, we focus our attention
on the latter regime. Given the series of $n$ beads representing the jet, we define the observable:

\begin{equation}
R_{i}=\frac{l_{step}}{l_{i}}
\end{equation}

where $l_{i}$ is the mutual distance between $i$ and $i+1$ elements, and
$l_{step}$ the length step.
We stress that $R\in(0,1]$, and, in particular,
$R_0$ is equal to 1 at the nozzle, since $l_{0}$ is imposed equal
to the length step $l_{step}$, which is used to discretize the jet
before stretching, and it works as a resolving power
indicator of the actual jet representation at a specific point of
the filament. Note that in the following, we use the notation $R(\lambda)$ instead of $R_{i}$ just for the
sake of simplicity, exploiting that, given a configuration, a unique value of $\lambda$ is associated to each $i-th$ bead,  
where the observable $R_{i}$ is computed. 
In Fig. \ref{Fig:resolving} we show
the mean value $<R\left(\lambda\right)>$ computed as function of the
curvilinear coordinate $\lambda$ over the entire stationary regime.
Here, we observe a quick decrease of $<R\left(\lambda\right)>$ for
the \textit{case I}. For instance, the resolving power $R$ is already
1/40 of its initial value at $\lambda$ equal to 0.1, and it continues
to decrease up to 1/200 of the initial $R$ value close the collector
at curvilinear coordinate $\lambda=1$. On the other hand, the resolving
power $R$ for the $case II$ is larger than 0.075 for any value of $\lambda$,
showing that the algorithm is efficient in preventing too coarse discretization
of the jet. Thus, we obtain the finer representation of the polymeric
filament displayed in Fig. \ref{Fig:representation-filament}, where
the beads belonging to the new discretization are colored in red,
and the grains of the old one are in blue.

\begin{figure}[H]
\begin{centering}
\includegraphics[scale=0.45]{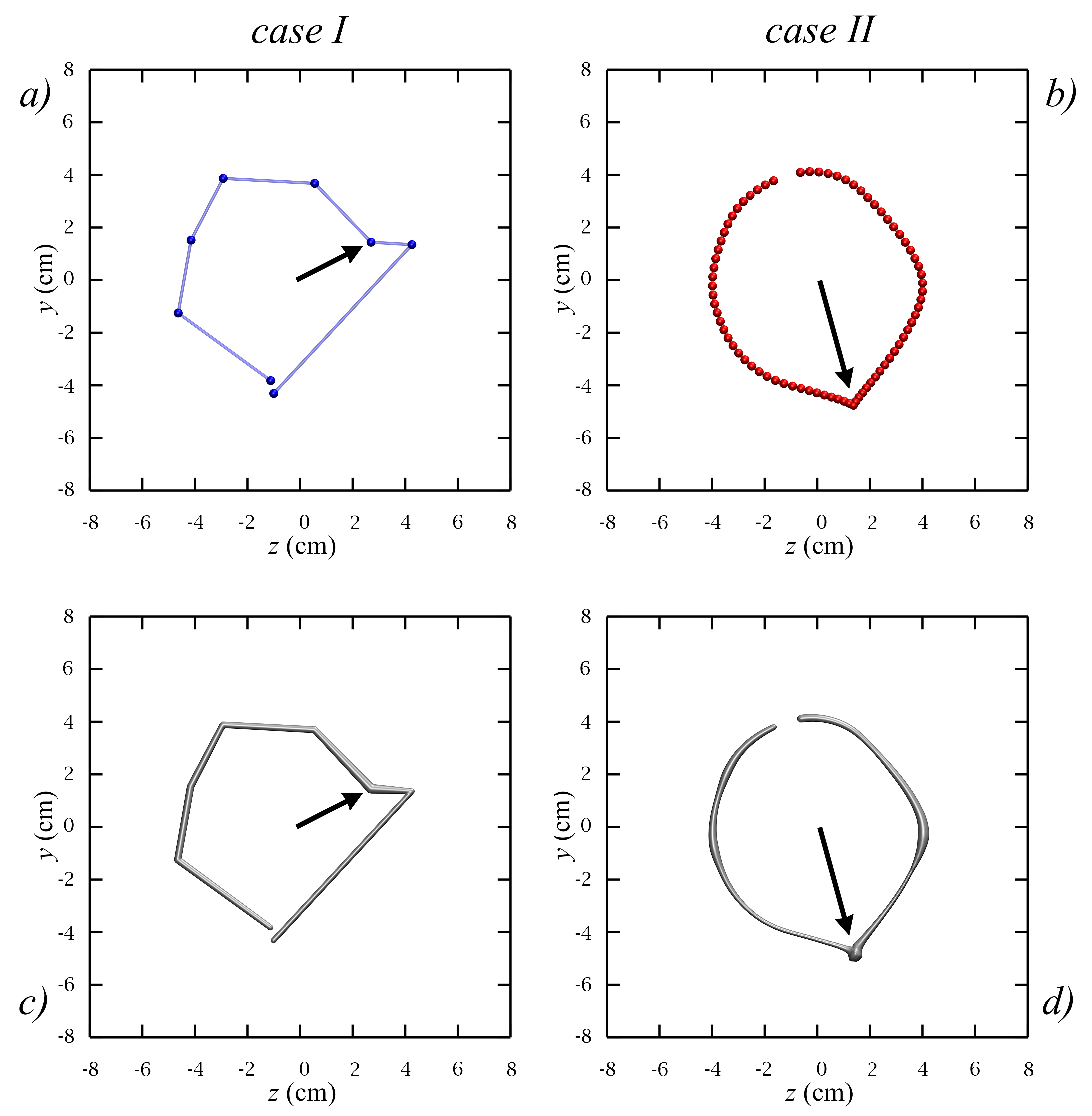} 
\par\end{centering}

\protect\caption{Snapshots of the nanofiber segments deposited on the collector for
the \textit{case I} on top left (\textit{a}) and for the \textit{case
II} in red dashed line on top right (\textit{b}). In black arrows
we highlight varicose defects along the filaments.
We also report the fiber radius profile for the \textit{case I} on
bottom left (\textit{c}), and for the \textit{case II} on bottom right
(\textit{d}). The cross section radius was multiplied by three orders
of magnitude in order to highlight varicosity.}

\label{Fig:collector} 
\end{figure}

The higher resolving power $R$ of \textit{case II} safeguards a finer
discrete representation of the jet which is a continuous object.
This is particularly useful in modeling varicosity \cite{yang2014crossover}.
As an example, we report an electrospinning simulation of
a varicose jet in order to test our refinement method in this
special framework. 
Here, varicose jet stands for a filament with
rapid fluctuations in its cross section radius.
The varicosity was artificially induced by inserting
at the nozzle a bead every ten beads with an extra 10\% of mass, while all
the other parameters are kept the same as in Tab. \ref{tab:simulation-param}.
This extra mass models a solid impurity, which triggers the varicosity
along its jet path from the nozzle towards the collector. In Fig. \ref{Fig:collector}
we report a short segment of the filament deposited at the collector
for the two cases. Here, we note that in \textit{case II} the mesh
refinement is capable of representing the fluctuation of the cross
section along the nanofiber. On the other hand, such information is
completely missed in the \textit{case I}, where the filament results
at lower resolution. This highlights the utility of the dynamic refinement
procedure, whenever our aim is to model nanofiber features of small
length-scale.

\begin{figure}[H]
\begin{centering}
\includegraphics[scale=0.70]{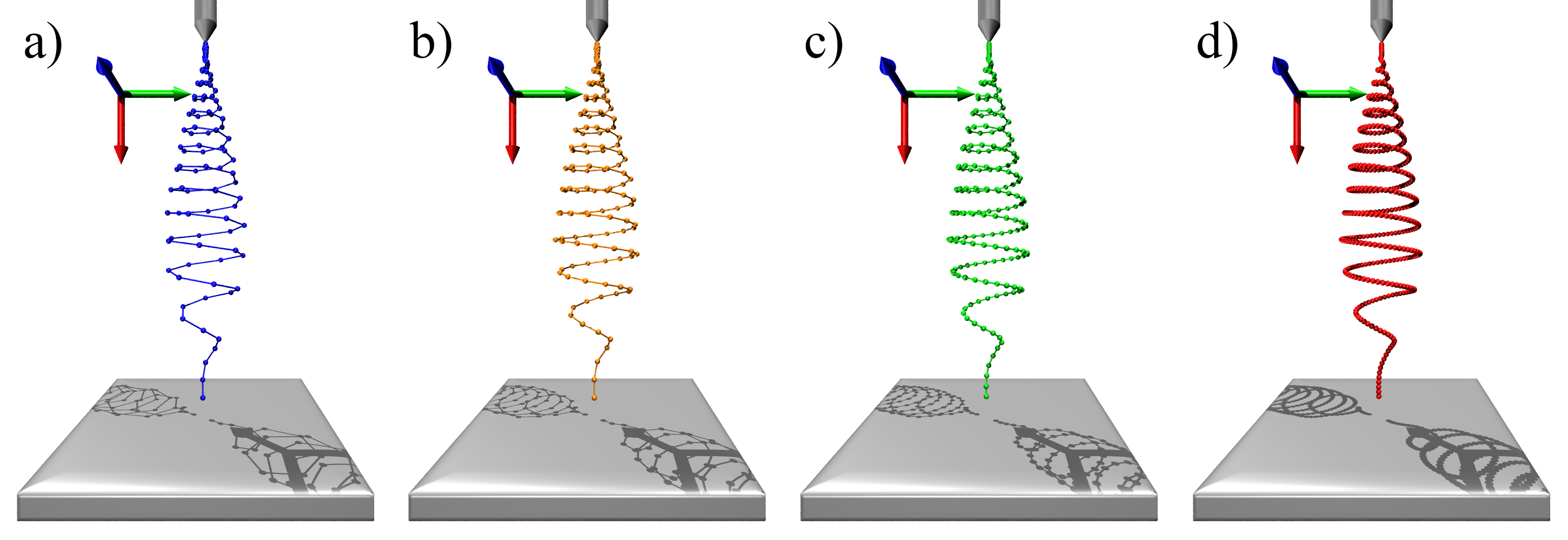} 
\par\end{centering}

\caption{Snapshots of the liquid jet taken at time $t$ equal to 0.009 seconds
for four different simulations with identical simulation parameters apart the value of $\tilde{l}$. Starting from left, denoted a) simulation without dynamic refinement activated,
b), c) and d) tree different simulations with dynamic refinement activated and $\tilde{l}$ equal to 0.2 cm, 0.4 cm and 0.8 cm,
respectively. Note in figure we have added two lights along the bisector of the plane $y-z$ in order to emphasize the depth of field.}

\label{Fig:resolution-comp4} 
\end{figure}

Next, we offer few comments on the parameter $\tilde{l}$ which defines
the resolution parameter used in the spline interpolation of the string.
It was observed by Yarin et al \cite{yarin2001taylor} that the discrete
element modeling of a fiber by 0-dimensional point particles (beads)
implies mathematical inconsistencies, as the discretization density
increases. Indeed, the modeled fiber tends to an infinitely thin,
one-dimensional continuous object, invalidating thus the discrete
bead model. The repulsive Coulomb force is the term incurring in such
issue, since the electrostatic force between two consecutive jet segments
of length $l_{step}$ and with their junction point at $\lambda_{p}$
value in curvilinear coordinate is computed by the double integral

\begin{equation}
\vec{\textbf{f}}_{1\rightarrow2}=\frac{1}{4\pi\varepsilon_{0}}{\displaystyle \int_{\lambda_{p}-l_{t}}^{\lambda_{p}}d\lambda_{1}\int_{\lambda_{p}}^{\lambda_{p}+l_{t}}d\lambda_{2}\frac{q_{l}^{2}}{\left|\vec{\textbf{r}}\left(\lambda_{2}\right)-\vec{\textbf{r}}\left(\lambda_{1}\right)\right|^{3}}}\left(\vec{\textbf{r}}\left(\lambda_{2}\right)-\vec{\textbf{r}}\left(\lambda_{1}\right)\right)\label{eq:Coulomb-int}
\end{equation}
where $\vec{\textbf{r}}\left(\lambda\right)$ denotes the position
vector along the string at curvilinear coordinate $\lambda$, and
$q_{l}$ is the linear charge density. It is easily seen that the integral diverges as the parameter $\tilde{l}$ tends to zero. In an
analytical study Yarin et al.\cite{yarin2001taylor} addressed this
issue by arguing that charges are not distributed on the centerline
of the fiber, but they lies rather on its outer shell of circular
shape. Thus, the problem is recovered by accounting for the actual
electrostatic form factors between two interacting circular sections
of a charged filament. In this context, several analytical approximations
were developed \cite{yarin2001taylor,kowalewski2005experiments,kowalewski2009modeling}
for computing the local electric force acting between two contiguous
rings representing jet elements. 
We recommend to use one of the mentioned
analytical approximations, if really low values of $\tilde{l}$ are
employed during the dynamics refinement procedure. However, a reasonable
value of $\tilde{l}$ can be assessed in order to maintain a fine
jet representation and avoiding, at the same time, mathematical inconsistencies.
In particular, the Coulomb integral for the electrostatic repulsion
between two interacting rings can be numerically computed and compared
with Eq. \ref{eq:Coulomb-int} in order to estimate the error introduced
by the centerline approximation at the given value of $\tilde{l}$.
A conservative value of $\tilde{l}$ larger than the length step $l_{step}$
introduces a negligible error due to the centerline approximation.
In this work, we impose a value of $\tilde{l}$ which is an order
of magnitude larger than length step $l_{step}$.
In order to justify the last statement, we report in Fig \ref{Fig:resolution-comp4}
snapshots of several simulations performed with three distinct
values of $\tilde{l}$ equal to 0.2 cm, 0.4 cm, 0.8 cm, and
a fourth simulation without the dynamic refinement activated at time $t$ equal to 0.009 seconds,
just 0.001 seconds before the jet touches the ground.
Here, it is evident as the jet topology is substantially equivalent,
confirming that possible errors introduced by
our approximation for the Coulomb force of Eq \ref{eq:force-coul} 
at our operative values of $\tilde{l}$ are negligible.

Furthermore, it is worth to stress that a small value of $\tilde{l}$
also implies a larger computational cost of the simulation. Indeed,
the computation of the repulsive Coulomb force is a $n$-body problem,
which scales as $\mathcal{O}\left(n^{2}\right)$ with $n$ denoting
the number of discrete elements. As example, we report in Tab. \ref{Tab:cpu-time}
the CPU wall-clock time of the three electrospinning simulations,
with the dynamic refinement procedure and $\tilde{l}$ equal to 0.2 cm, 0.4 cm, and 0.8 cm, 
previously mentioned. Along with
these data, the computational cost of a simulation without dynamic
refinement is also shown. For the sake of completeness, we also report
in Tab. \ref{Tab:cpu-time} the parallel efficiency $\eta$, which
is defined as $\eta=T_{s}/\left(T_{p}*n_{proc}\right)$ with $T_{s}$
and $T_{p}$ the CPU wall-clock time for the same job executed in
serial and parallel mode, and $n_{proc}$ the number of CPUs involved
in the parallel run. Here, we note that the increase of effective
CPU time is dependent on the average number of discrete elements $n$,
as expected. In particular, we note that the simulation with dynamic
refinement implies a higher CPU cost than a classical simulation (\textit{case
I}). In Ref. \cite{kowalewski2009modeling}, Kowalewski et al. addressed
the problem by using the hierarchical force calculation algorithm
(treecode) for computing the Coulomb force, which complexity scales
as $\mathcal{O}\left(n\,\log n\right)$ \cite{barnes1986hierarchical}.
The fast multiple method (FMM) \cite{greengard1987fast} can be also
used to handle long-range interactions with a high computational efficiency,
which theoretically achieves $\mathcal{O}\left(n\right)$ operation
count. The treecode and FMM are the best candidate to overcome this
issue, and its implementation in our dynamic refinement code will
be considered in future works.

\section{Conclusions }

We have presented a dynamic mesh refinement method for addressing
low resolution problems in discrete element models of jet hydrodynamics.
In particular, we have shown by practical examples that the proposed
algorithm is able to recover a finer representation by enforcing a
uniform arc-length discretization of a fluid jet at constant time
interval, before the information describing the modeled object is
scattered downstream. Furthermore, the preserved jet representation is
able to model short-range phenomena, such as varicosity,
providing more realistic simulations. In this work, we have described
the basic structure of the algorithm and the main steps for its implementation
in discrete element models of the electrospinning process. In addition,
we have discussed the effect of low values of the resolution parameter
$\tilde{l}$ on discrete element models and on the relative computational
costs, so as to provide general guidelines for using the dynamics
refinement procedure in electrospinning models.

In summary, the presented algorithm may be used for investigating
complex short-range phenomena of unidimensional liquid jets as well
as to improve the accuracy of discrete element models in representing
electro-hydrodynamic processes. Several existing models might benefit
of dynamic refinement in order to better support experiments, and
provide useful insights for enhancing the efficiency of several processes
involving jet hydrodynamics.

\section*{Acknowledgments}

The research leading to these results has received funding from the
European Research Council under the European Union's Seventh Framework
Programme (FP/2007-2013)/ERC Grant Agreement n. 306357 (\textquotedbl{}NANO-JETS\textquotedbl{}).

\newpage{}

\section*{Tables}

\begin{table}[H]
\begin{centering}
\begin{tabular}{cccccccccc}
\hline 
$\rho$  & $\rho_{q}$  & $a_{0}$  & $\upsilon_{s}$  & $\alpha$  & $\mu$  & $G$  & $V_{0}$  & $\omega$  & $A$\tabularnewline
($\text{kg}/\text{m}^{3}$)  & ($\text{C}/\text{L}$)  & ($\text{cm}$)  & ($\text{cm/s}$)  & (mN/m)  & (Pa$\cdot$s)  & (Pa)  & (kV)  & ($\text{s}^{-1}$)  & ($\text{cm}$)\tabularnewline
\hline 
\hline 
840  & $2.8\cdot10^{-7}$  & $5\cdot10^{-3}$  & 0.28  & 21.1  & 2.0  & $5\cdot10^{4}$  & 9.0  & $10^{4}$  & $10^{-3}$\tabularnewline
\hline 
\end{tabular}
\par\end{centering}

\protect\protect\protect\caption{Simulation parameters for the simulations of PVP solution electrified jets. The headings
used are as follows: $\rho$: density, $\rho_{q}$: charge density,
$a_{0}$: fiber radius at the nozzle, $\upsilon_{s}$: initial fluid
velocity at the nozzle, $\alpha$: surface tension, $\mu$: viscosity,
$G$: elastic modulus, $V_{0}$: applied voltage bias, $\omega$:
frequency of perturbation at the nozzle, $A$: amplitude of perturbation
at the nozzle.}

\label{tab:simulation-param} 
\end{table}

\begin{table}[H]
\begin{centering}
\begin{tabular}{cccccc}
\hline 
Dynamic Refinement  & $\tilde{l}$  & \# of beads  & \# of CPUs  & CPU wall clock time  & $\eta${*}\tabularnewline
 & ($\text{cm}$)  &  &  & ($\text{s}$)  & \tabularnewline
\hline 
\hline 
no  &  & 150  & 24  & 8787  & 0.26\tabularnewline
yes  & 0.8  & 267  & 24  & 12718  & 0.44\tabularnewline
yes  & 0.4  & 509  & 24  & 24726  & 0.62\tabularnewline
yes  & 0.2  & 1123  & 24  & 75793  & 0.80\tabularnewline
\hline 
\end{tabular}
\par\end{centering}

\protect\protect\caption{We report the CPU wall-clock time in seconds which is needed to run
several simulations with and without dynamic refinement at different
values of $\tilde{l}$. We report also the average number of beads
used to discretize the jet. The benchmark was executed on a node of
2x12 core processors made of 2.4 GHz Intel Ivy Bridge cores. {*}It
is worth to stress that the parallel efficiency $\eta$ increases
with the number of beads, since the communication latency cost plays
a larger role as the simulation involves a small number of beads,
and the computational work is consequently not well distributed over
all the CPU cores.}

\label{Tab:cpu-time} 
\end{table}

\newpage{}

\section*{Figures}

\end{document}